\documentclass[a4paper,11pt]{article}
\usepackage{pos}

\title{Theoretical predictions to differential cross sections and decay rates from the loop-tree duality}
\ShortTitle{Theoretical predictions from the loop-tree duality}

\author*[a]{David F. Renter\'{\i}a-Estrada}
\affiliation[a]{Instituto de F\'{\i}sica Corpuscular, Universitat de Val\`encia - Consejo Superior de Investigaciones Cient\'{\i}ficas.\\
Parc Cient\'{\i}fic, Carrer del Catedr\`atic Jos\'e Beltr\'an Mart\'{\i}nez, 2, 46980 Paterna, Valencia, Spain}

\emailAdd{david.renteria@ific.uv.es}
\abstract{Understanding the cancellation of ultraviolet and infrared singularities in perturbative quantum field theory is of central importance for the development and automation of various theoretical tools that make accurate predictions for observables at high-energy colliders. The loop-tree duality aims to find an efficient solution by treating loop and tree-level contributions under the same foot to achieve a local cancellation of singularities at the integrand level, and thus avoiding dimensional regularisation. In this talk, we exploit the causal properties of vacuum amplitudes in the loop-tree duality representation to present different applications to physical processes at higher orders.}

\FullConference{%
  42nd International Conference on High Energy Physics (ICHEP2024)\\
  18-24 July 2024\\
  Prague, Czech Republic
}

\begin{document}
\maketitle
\section{Introduction}
The Loop-Tree Duality (LTD) framework \cite{Catani:2008xa,Aguilera-Verdugo:2020set} was developed to simplify the evaluation of scattering amplitudes in Quantum Field Theory (QFT) by facilitating the calculation of loop integrals. The core idea behind this method is to transform loop-momentum diagrams into a sum of tree-level diagrams, reducing the four-dimensional Minkowski integration space to a three-dimensional Euclidean phase-space. Scattering amplitudes, however, are mathematical objects that present significant challenges, particularly at higher orders where singularities of infrared (IR) and ultraviolet (UV) type emerge. These singularities are typically addressed using the dimensional regularization (DREG) method. As these singularites are the consequence of fixing the number of external particles in scattering amplitudes, we propose an alternative methodology based on vacuum amplitudes, where external particles are effectively removed from consideration. This work builds upon the articles \cite{Ramirez-Uribe:2024rjg,LTD:2024yrb,deLejarza:2024scm} where we settle the basis of the LTD Causal Unitarity (CU) formalism. With this brief presentation,  we illustrate how vacuum amplitudes {are the optimal building block} to compute higher-order corrections to physical observables. This framework exploits the LTD causal representation to achieve a natural cancellation of singularities, directly at integrand-level. 

%%%%%%%%%%%%%%%%%%%%%%%%%%%%%%%%%%%%%%%%%%%%%%%%%%%%%%%%%%%%%%%%%%%%%%%%%%%%%%%%%%%%
%%%%%%%%%%%%%%%%%%%%%%%%%%%%%%%%%%%%%%%%%%%%%%%%%%%%%%%%%%%%%%%%%%%%%%%%%%%%%%%%%%%%
\section{Fundamental concepts}
In particle physics, it is often necessary to compute loop Feynman integrals, which can be quite difficult to perform: this is particularly true when multiple external particles are present. To simplify this process, we use the LTD formalism: it allows us to express the loop diagram as tree-level objects integrated in a phase-space with additionally real particles. Specifically, at one loop\cite{JesusAguilera-Verdugo:2020fsn},
\begin{eqnarray}
    {\cal A}_{\rm D}^{(\Lambda)} = \int_{C_L}\prod_{i=1}^{n}G_F(q_i)
    = -2\pi\imath\sum{\rm Res}\left(\prod_{i=1}^{n}G_F(q_i), {\rm Im}(\eta \cdot q_i) < 0\right)\,.
\end{eqnarray}
In this context, \( G_F(q_i) \) represents the Feynman propagator of a single particle with four-momentum~\( q_i \). The condition \( \text{Im}(\eta \cdot q_i) < 0 \) selects the poles with negative imaginary components. The time-like vector $\eta$ can be used is arbitrary. If $\eta^\mu = (1, \mathbf{0})$, the energy components of the loop momenta are integrated out. A multiloop vacuum amplitude in the LTD representation \cite{Ramirez-Uribe:2024rjg} is defined over the Euclidean space of the spacial components of the loop momenta  \( \{\boldsymbol{\ell}_j\}_{j=1, \ldots, \Lambda} \), i.e.
\begin{align}
    {\cal A}^{(\Lambda)} = \int_{\boldsymbol{\ell}_1\,\ldots\, \boldsymbol{\ell}_\Lambda}{\cal A}_{\rm D}^{(\Lambda)}\,, \quad \int_{\boldsymbol{ \ell_j}} =  \mu^{4-d}\int\frac{d^{d-1}\ell_j}{(2\pi)^{d-1}}\,,
\end{align}
in \( d \)-dimensional space-time, and  \( \mu \) represents the dimensional regularisation (DREG) scale. Considering a vacuum amplitude in the LTD causal representation, the Feynman propagators are replaced by causal propagators, which can be generically written as:
\begin{eqnarray}
    \frac{1}{\lambda_{i_1\,i_2\,\ldots\,i_n}} = \frac{1}{\sum_{s=1}^{n}q_{i_s,0}^{(+)}}\,,
\end{eqnarray}
with \( q_{i_s,0}^{(+)} = \sqrt{\mathbf{q}_{i_s}^2+m_{i_s}^2-\imath\,0} \) the on-shell energy component where \( \mathbf{q}_{i_s} \) represents the spatial components of \( q_{i_s} \) and \( m_{i_s} \) is the mass of the propagating particle. The term \( \imath 0 \) arises from the original infinitesimal complex prescription of the Feynman propagator. In terms of causal cuts, each causal propagator involves a set of internal particles that divide the amplitude into two sub-amplitudes, with the momentum flow of all particles in the set aligned in the same direction. In terms of graph representations, two causal propagators are compatible if the common particles are aligned in the same direction, forming a directed acyclic graph (DAG) \cite{Sborlini:2021owe,Ramirez-Uribe:2021ubp,Clemente:2022nll,Ramirez-Uribe:2024wua}. The on-shell energies are, by definition, positive quantities  in the limit in which the complex prescription vanishes. As a result, the causal propagators become singular, \( \lambda_{i_1\,i_2\,\ldots\,i_n} \to 0 \),
only when the all the on-shell energies tend to an infinitesimally small value, which is an nonphysical configuration. Consequently, the vacuum amplitude cannot generate any soft, collinear or threshold singularities, allowing only ultraviolet (UV) singularities. Furthermore, in the LTD causal representation, this property holds at integrand-level, which makes this formalism particularly effective to avoid spurious singularities and numerical instabilities.

%%%%%%%%%%%%%%%%%%%%%%%%%%%%%%%%%%%%%%%%%%%%%%%%%%%%%%%%%%%%%%%%%%%%%%%%%%%%%%%%%%%%
%%%%%%%%%%%%%%%%%%%%%%%%%%%%%%%%%%%%%%%%%%%%%%%%%%%%%%%%%%%%%%%%%%%%%%%%%%%%%%%%%%%%
\section{Differential observables from Vacuum Amplitudes}
The idea of eliminating all external particles stems from the hypothesis that all quantum fluctuations contributing to scattering or decay processes at high energies are contained within the vacuum amplitude, where the final and initial states correspond to selected cuts in the causal LTD representation: this is the cornerstone of Causal Unitarity. We propose the following master formula for the contributions to the differential cross-section at N$^k$LO order
\begin{eqnarray}
    d\sigma_{{\rm N}^k{\rm LO}} = \frac{d\Lambda}{2s} \,  
\sum_{(i_1\cdots i_n a b) \in \Sigma} {\cal A}_{\rm D}^{(\Lambda, {\rm R})}(i_1\cdots i_n a b) \, 
{\cal O}_{i_1\cdots i_n} \, \tilde \Delta_{i_1\cdots i_n \bar a \bar b}~,
\end{eqnarray}
along with the corresponding integration measure 
\begin{eqnarray}
    d\Lambda = \prod_{j=1}^{\Lambda-2} d\Phi_{\boldsymbol{\ell}_j} = \prod_{j=1}^{\Lambda-2} \mu^{4-d} \frac{d^{d-1} \boldsymbol{\ell}_j}{(2\pi)^{d-1}}~\,.
\end{eqnarray}
The  upper limit $\Lambda-2$ arises because two of the loop three-momenta are determined by the initial-state conditions.
\begin{figure}
    \centering
    \includegraphics[width=0.25\linewidth]{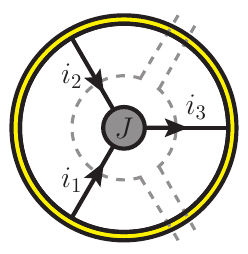}
    \caption{Graphical interpretation of the local cancellation of double-collinear final-state contributions in the dual vacuum amplitude.}
    \label{fig:kernelfinalstate}
\end{figure}
Energy conservation is enforced through the Dirac delta function $\tilde \Delta_{i_1\cdots i_n \bar a\bar b} =  2\pi \,  \delta(\lambda_{i_1\cdots i_n \bar a\bar b})~$. On the other hand, the observable to be calculated is denoted by \({\cal O}_{i_1\cdots i_n}\), where the total cross-section and decay rate are recovered when ${\cal O}_{i_1\cdots i_n} = \boldsymbol{1}$. The renormalised phase-space residues are obtained from  the vacuum amplitude $ {\cal A}_{\rm D}^{(\Lambda)}$, 
\begin{eqnarray}
    {\cal A}_{\rm D}^{(\Lambda,{\rm R})}(i_1 \cdots i_n a b) = {\rm Res} \left(\frac{x_{ab}}{2} \, {\cal A}_{\rm D}^{(\Lambda)},  \lambda_{i_1\cdots i_n a b}\right) - {\cal A}_{\rm UV/C}^{(\Lambda)}(i_1 \cdots i_n a b)\,.
\end{eqnarray}
The counterterm \( {\cal A}_{\rm UV/C}^{(\Lambda)} \) is responsible for implementing a local ultraviolet (UV) renormalization and facilitating the local subtraction of collinear singularities in the initial state. A notable property of the vacuum amplitude is its inherent ability to naturally remove collinear singularities. Figure \ref{fig:kernelfinalstate} provides a diagrammatic interpretation of the local cancellation of double-collinear singularities in the final state. The causal amplitude representation is proportional to two causal propagators, i.e.
\begin{eqnarray}
    {\cal A}_{\rm D}^{(\Lambda)} \sim \frac{1}{\lambda_{i_1 i_2 \cdots ab} \, \lambda_{i_3 \cdots ab}}~.
\end{eqnarray}
In this framework, $a$ and $b$  represent the initial state. By evaluating the phase-space residues, we arrive at the following result:
%\begin{eqnarray}
%{\cal A}_{\rm D}^{(\Lambda)}(i_3 \cdots ab) = {\frac{1}{\lambda_{i_1i_2\bar i_3}}}~, \\ 
%{\cal A}_{\rm D}^{(\Lambda)}(i_1 i_2 \cdots ab) = { - \, \,  \frac{1}{\lambda_{i_1i_2\bar i_3}}}~.
%\end{eqnarray}
\begin{eqnarray}
{\cal A}_{\rm D}^{(\Lambda)}(i_3 \cdots ab) = {\frac{1}{\lambda_{i_1i_2\bar i_3}}}~, \qquad 
{\cal A}_{\rm D}^{(\Lambda)}(i_1 i_2 \cdots ab) = { - \, \,  \frac{1}{\lambda_{i_1i_2\bar i_3}}}~.
\end{eqnarray}
This leads to two divergent contributions in the collinear limit $\lambda_{i_1i_2\bar i_3} \to 0$, but with opposite signs. This implies that,  the sum of both collinear singularities cancels, i.e.
\begin{eqnarray}
    \lim_{\lambda_{i_1 i_2 \bar i_3}\to 0} \left( {\cal A}_{\rm D}^{(\Lambda)}(i_1 i_2 \cdots a b) \, \tilde \Delta_{i_1 i_2 \cdots \bar a \bar b}  + {\cal A}_{\rm D}^{(\Lambda)}{(i_3 \cdots a b)} \, \tilde \Delta_{i_3 \cdots \bar a \bar b} \right) = {\cal O} (\lambda_{i_1 i_2 \bar i_3}^0)~.
\end{eqnarray}
The local cancellation is consistent with energy conservation: \(\lim_{\lambda_{i_1i_2\bar{i}_3}\to 0} \tilde \Delta_{i_1 i_2 \cdots \bar a \bar b}  = \tilde\Delta_{i_3 \cdots \bar a \bar b} \,.\) As a proof of concept at next-to-leading order (NLO), we have calculated the decay rate of the $H \to q\Bar{q}(g)$ process. This observable can be derived from the three-loop vacuum amplitude, which acts as a \emph{kernel dual amplitude} at NLO, 
\begin{figure}
    \centering
    \includegraphics[width=0.5\linewidth]{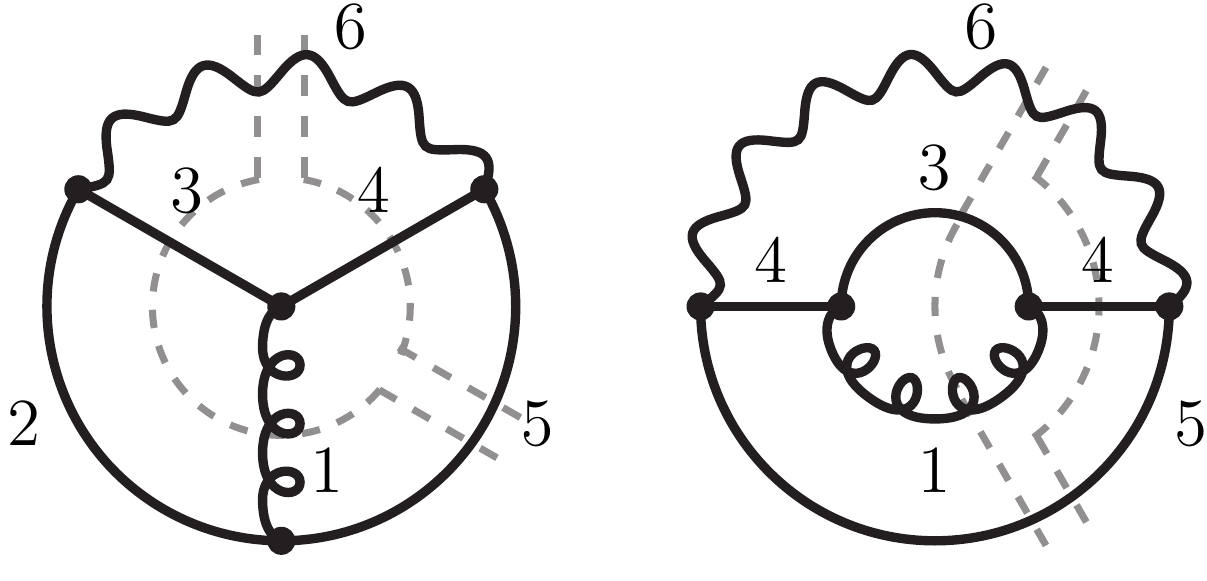}
    \caption{The three-loop vacuum diagrams that generate \(H\to q\bar{q}(g)\) at NLO. The gray dashed lines indicate phase-space residues.}
    \label{fig:VAnlo}
\end{figure} 
\begin{eqnarray}
d\Gamma^{(1)}_{H\to q\bar q} &= \frac{d\Phi_{\boldsymbol{\ell}_1\boldsymbol{\ell}_2}}{2\sqrt{s}} \, \bigg[
\Big( {\cal A}_{\rm D}^{(3,H,{\rm R})}(456) \, \tilde \Delta_{45\bar 6} + {\cal A}_{\rm D}^{(3,H)}(1356) \, \tilde \Delta_{135\bar 6} \Big)  + (5\leftrightarrow 2,~ 4\leftrightarrow 3) \bigg]~.
\end{eqnarray}
Figure \ref{fig:VAnlo} displays the corresponding kernel vacuum amplitude: the details of the calculation are exposed in Ref.~\cite{LTD:2024yrb}. In this case, the phase-space residues correspond to contributions with three (virtual contribution) or four external particles (real contribution). 
\begin{figure}
    \centering
    \includegraphics[width=0.45\linewidth]{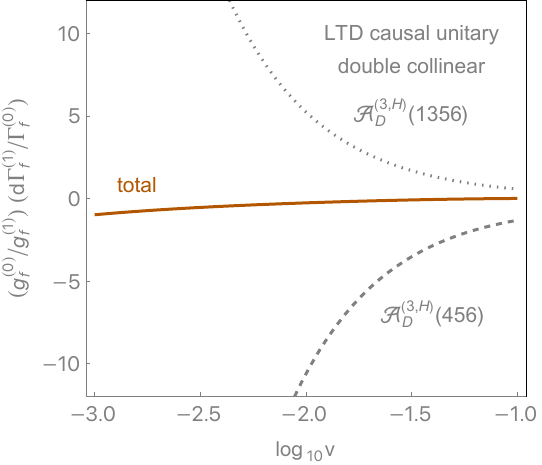}
    \caption{At NLO, collinear singularities locally cancel between phase-space residues with differing numbers of final-state particles.}
    \label{fig:LTD-contribution}
\end{figure}

The integrand behaviour of the phase-space residues is shown in Fig. \ref{fig:LTD-contribution}, where \( {\cal A}_{\rm D}^{(3,H,{\rm R})}(456) \) and \( {\cal A}_{\rm D}^{(3,H,{\rm R})}(1356) \) exhibit singularities in the  collinear configuration: remarkably, the sum of both contributions remains flat, i.e. \emph{divergences cancel locally}. Additionally, in Ref. \cite{LTD:2024yrb}, we performed a numerical implementation of LTDCU at NLO, where the results exhibit a dependence on the renormalization scale \( \mu_{UV} \). On the other hand, in Ref.~\cite{deLejarza:2024scm}, the  decay rate was calculated numerically using a novel quantum integration algorithm known as Quantum Fourier Interactive Amplitude Estimation (QFIAE)~\cite{deLejarza:2023qxk}. In Ref.~\cite{deLejarza:2024scm}, we present  results conducted on a quantum simulator and on a real quantum device, IBMQ superconducting {\it ibmq\_mumbai}. The results in the quantum simulator showed very good agreement with respect to the analytical predictions within the DREG framework. Furthermore, only tiny discrepancies due to the noise introduced by the quantum hardware were observed.
 
%%%%%%%%%%%%%%%%%%%%%%%%%%%%%%%%%%%%%%%%%%%%%%%%%%%%%%%%%%%%%%%%%%%%%%%%%%%%%%%%%%%%
%%%%%%%%%%%%%%%%%%%%%%%%%%%%%%%%%%%%%%%%%%%%%%%%%%%%%%%%%%%%%%%%%%%%%%%%%%%%%%%%%%%%
\section{Conclusions}
In this work, we have highlighted one of the key advantages of the LTD Causal Unitarity framework, based on cutting a kernel vacuum amplitude for calculating scattering amplitudes at higer-orders. Using the example of the decay rate of the process $H\to q\bar{q}(g)$ at NLO, we successfully demonstrated that collinear singularities are locally canceled. Additionally, we performed the numerical computation of the integrals for the decay rate in a quantum implementation using the QFIAE algorithm. The results obtained from the quantum simulator closely matched those obtained through numerical computation with DREG, while the results using hardware showed deviations only due to quantum device noise: in the future, we expect error mitigation techniques to provide more accurate results.
%%%%%%%%%%%%%%%%%%%%%%%%%%%%%%%%%%%%%%%%%%%%%%%%%%%%%%%%%%%%%%%%%%%%%%%%%%%%%%%%%%%%
%%%%%%%%%%%%%%%%%%%%%%%%%%%%%%%%%%%%%%%%%%%%%%%%%%%%%%%%%%%%%%%%%%%%%%%%%%%%%%%%%%%%
\section{Acknowledgements}
%A la Grisolia por pagarme mi sueldito. 
I would like to thank Germán F. R. Sborlini, Selomit Ramírez-Uribe, Prasanna K. Dhani, and Germán Rodrigo for reading this manuscript, and extend special thanks for the support provided during this talk. This work is supported by the Spanish Government - Agencia Estatal de Investigaci\'on (MCIN/ AEI/10.13039/501100011033) Grant No. PID2020-114473GB-I00, and Generalitat Valenciana, Grants No. PROMETEO/2021/071 and ASFAE/2022/009 (Planes Complementarios
de I+D+i, NextGenerationEU). This work is also supported by the Ministry of Economic Affairs and Digital Transformation of the Spanish Government and NextGenerationEU through the Quantum Spain project, and by CSIC Interdisciplinary Thematic Platform on Quantum Technologies (PTI-QTEP+). DFRE is supported by Generalitat Valenciana, Grant No. CIGRIS/2022/145.
%%%%%%%%%%%%%%%%%%%%%%%%%%%%%%%%%%%%%%%%%%%%%%%%%%%%%%%%%%%%%%%%%%%%%%%%%%%%%%%%%%%%
%%%%%%%%%%%%%%%%%%%%%%%%%%%%%%%%%%%%%%%%%%%%%%%%%%%%%%%%%%%%%%%%%%%%%%%%%%%%%%%%%%%%
%\providecommand{\href}[2]{#2}\begingroup\raggedright\begin{thebibliography}{10}

%\end{thebibliography}\endgroup

%\bibliographystyle{JHEP}
%\bibliography{refs}

%%%%%%%%%%%%%%%%%%%%%%%%%%%%%%%%%%%%%%%%%%%%%%%%%%%%%%%%%%%%%%%%%%%%%%%%%%%%%%%%%%%%
%%%%%%%%%%%%%%%%%%%%%%%%%%%%%%%%%%%%%%%%%%%%%%%%%%%%%%%%%%%%%%%%%%%%%%%%%%%%%%%%%%%%
\end{document}